\documentclass[pra,aps,twocolumn,floatfix, superscriptaddress,a4paper,accepted=2023-01-11]{quantumarticle}
\pdfoutput=1
\usepackage{graphicx}
\graphicspath{ {./images/} }
\usepackage[table]{xcolor}
\usepackage{amsmath}

\usepackage{amssymb}
\usepackage{bbold}
\usepackage{braket}
\usepackage[font=small, justification=raggedright]{caption}
\usepackage[font=small, justification=raggedright]{subcaption}
\usepackage{amsthm}
\usepackage{setspace}
\usepackage[T1]{fontenc}

\usepackage[numbers]{natbib}

\usepackage{hyperref,color}
\definecolor{darkblue}{RGB}{0,0,149}
\hypersetup{colorlinks=true,linkcolor=darkblue,filecolor=darkblue,urlcolor=darkblue,citecolor=darkblue,linktocpage=true}
\usepackage{url}

\newcommand{\rev}[1]{#1}
\newcommand{\fedit}[1]{#1}

\newcommand{\cedit}[1]{{#1}}

\newcommand{\medit}[1]{{#1}}

\begin{document}

\title{Gravitational time dilation as a resource in quantum sensing}

\author{Carlo Cepollaro}
\affiliation{
 Quantum Technology Lab, Dipartimento di Fisica Aldo Pontremoli, Universit\`a degli Studi di Milano, I-20133 Milano, Italy
}%
\affiliation{%
	Vienna Center for Quantum Science and Technology (VCQ), Faculty of Physics, University of Vienna, Boltzmanngasse 5, A-1090 Vienna, Austria
}%
\affiliation{%
	Institute of Quantum Optics and Quantum Information (IQOQI), Austrian Academy of Sciences, Boltzmanngasse 3, A-1090 Vienna, Austria
}%
\email{carlo.cepollaro@oeaw.ac.at}

\author{Flaminia Giacomini}
\affiliation{%
    Perimeter Institute for Theoretical Physics, 31 Caroline St. N, Waterloo, Ontario, N2L 2Y5, Canada
}%
\email{fgiacomini@perimeterinstitute.ca}

\author{Matteo G. A. Paris}
\affiliation{Quantum Technology Lab, Dipartimento di Fisica Aldo Pontremoli, Universit\`a degli Studi di Milano, I-20133 Milano, Italy
}%
\affiliation{INFN, Sezione di Milano, I-20133 Milano, Italy}
\email{matteo.paris@fisica.unimi.it}

\begin{abstract}
\medit{Atomic clock interferometers are a valuable tool to test the interface between quantum theory and gravity, in particular via the measurement of gravitational time dilation in the quantum regime. Here, we investigate whether
gravitational time dilation may be also used as a resource in quantum information theory. In particular, we show that for a freely falling interferometer and for a Mach-Zehnder interferometer, the gravitational time dilation may enhance the precision in estimating the gravitational acceleration for long interferometric times. To this aim, the interferometric measurements should be performed on both the path and the clock 
degrees of freedom.}
\end{abstract}

\maketitle

\section{Introduction}\label{sec:introduction}

\medit{Quantum mechanics changed the way we look at the physical world, and in the last two decades, quantum features of physical systems have also become a resource in different branches of technology \cite{Yin2020,Green2021}. In particular, when metrology met quantum mechanics, an entire class of novel features were made available to improve precision of physical measurements, and to conceive novel quantum-enhanced protocols to characterize signals and devices
\cite{qm1,qm2,qm3}. Relativity too changed the paradigms of physics, and found relevant technological applications \cite{gps}. A question thus arises of whether relativistic and quantum mechanical features may be jointly exploited to improve precision of physical measurements.}

\medit{In this paper, we follow this idea and prove that a paradigmatic relativistic feature, \fedit{gravitational} time dilation, may indeed represent a resource that can be used together with quantum superpositions to improve precision in the estimation of the gravitational constant, or of its variations.}

\fedit{Concretely, this enhancement can be realised in an experiment by employing quantum clocks. Quantum clocks have been studied as a tool to probe the interface between quantum theory and general relativity~\cite{greenberger, lammerzahl, mueller, zych2016, pikovski_TimeDilation, mueller2019, roura, roura2020, dipumpo}. For instance, using a quantum clock in an interferometric setup leads to a loss of visibility due to general relativistic proper time~\cite{zych_interferometric}. From a foundational perspective, quantum clocks have been studied in connection with gravitational decoherence~\cite{pikovski}, shown to set limits to the measurability of spacetime intervals~\cite{saleckerwigner, castro-ruiz, altamirano}, and to give rise to a relative time-localisation of events in the presence of indefinite causal order~\cite{zych2019bell, castro2020quantum}. In addition, quantum clocks have been used to propose generalisations of special and general relativistic proper time~\cite{khandewal, smith2020quantum, grochowski, giacomini_SQRF,Tobar2022}, as well as a quantum version of the Einstein Equivalence Principle~\cite{,sev17,sev18,zych_QEEP, cepollaro2021}}.

\rev{More generally, the fascinating interface between relativistic effect and quantum metrology has been explored in the last decade~\cite{Ahmadi2014Quantum,Yao2014Quantum}, e.g. to improve quantum measurement technologies~\cite{Ahmadi2014Relativistic,Hosler2013Parameter} and to characterize noninertial effects~\cite{Aspachs2010Optimal,Wang2014Quantum}.}

\rev{In this paper, we focus on using relativistic features to improve the precision of gravitational measurements. In particular, we prove that gravitational time dilation may be used to enhance the estimation of the gravitational constant. To the best of our knowledge, the metrological enhancement due to quantum clocks in an interferometric setup has not been studied before.}

\medit{The paper is structured as follows. In the next section, we establish notation and address the dynamics of a massive quantum clock in a gravitational field. In Section \ref{s:ffi}, we study the dynamics of a quantum clock when it traverses a freely falling interferometer, whereas in Section \ref{s:mzi} we address the specific case of an atom propagating in a Mach-Zehnder interferometer. Section \ref{s:out} closes the paper with some concluding remarks.}

\section{Massive quantum clock in a gravitational field} In this section, we review the Hamiltonian \fedit{formulation of the dynamics of an atomic clock in the gravitational field of the Earth}. \fedit{Following \cite{pikovski}, we begin with} a classical description, \fedit{which we then quantize}.

\fedit{A quantum clock is usually modelled as an atom with internal degrees of freedom. The full dynamics of a composite quantum particle in an arbitrary gravitational field is very complicated, but for the purposes of this work it is enough to consider the limit of weak gravitational field, non-relativistic velocities, and small accelerations. We treat the atom as point-like, with the relevant degrees of freedom, which are assigned a quantum state, being its center of mass and its internal structure (clock degrees of freedom). For a discussion of composite systems in a gravitational field, see Ref.~\cite{zych2019}.}

For a quantum clock moving vertically in the Earth gravitational field, and with the latter expressed in polar coordinates, the angular component of the metric does not contribute to the dynamics. Hence, the metric can be written in a 1+1 formulation, where to our order of approximation
\begin{equation}\label{eq:gravityfield}
   \begin{split}
        &g_{00}(x) = 1+2\frac{V_N(x)}{c^2}, \\
        &g_{01}(x) = g_{10}(x) = 0,\\
        &g_{11}(x) = -\left(1-2\frac{V_N(x)}{c^2}\right),
   \end{split}
\end{equation}
and $V_N(x)$ is the Newtonian potential. \fedit{Here, we have chosen the coordinate system of} an observer comoving with a laboratory \fedit{situated on the surface of the Earth}. In these coordinates, the relativistic dispersion relation \fedit{of the clock} reads
\begin{equation}\label{eq:dispersion_relation}
    p_\mu p^\mu = (E_{rest}/c)^2,
\end{equation}
where \fedit{$p_\mu$ is the momentum in $1+1$ dimensions} and $E_{rest}$ is the rest energy of the atom, \fedit{namely}
\begin{equation}
    E_{rest} = mc^2 + E_{int}.
\end{equation}
The split between mass and internal energy coincides with the choice of the zero-energy level for the internal energy, and it depends on the energy scale of the process considered: if some degrees of freedom do not evolve dynamically in a certain process, their energy is accounted for in the rest mass, as it happens for example with binding energies of the constituents of the atoms.

\fedit{Writing Eq.~\eqref{eq:dispersion_relation} explicitly in the weak gravitational field of Eq.~\eqref{eq:gravityfield} and solving for $p_0$ (the generator of the time translations in the coordinate time $t$), we obtain the Hamiltonian}
\begin{equation}
    H = c p_0 = \sqrt{g_{00}(-c^2 g^{ij} p_i p_j + E_{rest}^2)}.
\end{equation}
In the limit of weak gravitational field and non-relativistic velocities, we discard any term of the order $\frac{V_N(x) p^2}{m^2 c^4}$, obtaining
\begin{align}\label{eq:cl_hamiltonian}
    H = mc^2 & + \frac{p^2}{2m} + mV_N(x) + \notag \\ & + E_{int}\left(1+\frac{V_N(x)}{c^2}-\frac{p^2}{2m^2c^2}\right).
\end{align}
The corrections to the internal energy $E_{int}$ are the gravitational and special relativistic time dilation effects respectively. In fact, the time of the atomic clock is measured by synchronizing to the frequency of a photon emitted in an electronic transition (see e.g. Ref.~\cite{ludlow}), and the frequency is proportional to the energy difference between the electronic levels, which are modified \fedit{due to special and general relativistic time dilations}. Therefore, the frequency is
\begin{equation}\label{eq:omega_prime}
    \omega^\prime = \omega \left(1+\frac{V_N(x)}{c^2}-\frac{p^2}{2m^2c^2}\right),
\end{equation}
where $\omega = \frac{\Delta E}{\hbar}$ is the frequency measured in the locally inertial frame \fedit{comoving with} the atom. Therefore, \fedit{the time dilation} factor
\begin{equation}\label{eq:redshift_factor}
    \left(1+\frac{V_N(x)}{c^2}-\frac{p^2}{2m^2c^2}\right) \equiv \frac{d\tau}{dt},
\end{equation}
is the rate at which the proper time of the atom flows in the laboratory time.

This derivation may be repeated for different choices of the metric, as long as they are static, and the same results can also be found with a quantum field theory approach~\cite{pikovski}. \fedit{An equivalent derivation in the low-energy limit is obtained starting from} the Newtonian Hamiltonian $H_{N}= mc^2 + p^2/2m  + m\,V_N(x)$ by inserting the mass-energy equivalence $m\to m + E_{int}/c^2$.

The quantization of the Hamiltonian in Eq.~\eqref{eq:cl_hamiltonian} is straightforward and leads to
\begin{align}
\label{eq:q_hamiltonian}
    \hat H = mc^2 & + \frac{\hat p^2}{2m} +   mV_N(\hat x) + \notag \\  & + \hat H_{int}\left(1+\frac{V_N(\hat x)}{c^2} -\frac{\hat p^2}{2m^2c^2}\right).
\end{align}
\fedit{In the quantum case, the clock space is} a two-dimensional Hilbert space $\mathcal{H}_{int}$ \fedit{spanned by the} basis $\{\ket{0},\ket{1}\}$. The internal Hamiltonian is
\begin{equation}
        \hat H_{int} = E_0 \ket{0}\bra{0} + E_1 \ket{1}\bra{1}.
\end{equation}

\section{Freely Falling Interferometer}
\label{s:ffi}
\begin{figure}[h!]
    \includegraphics[width=0.95\columnwidth]{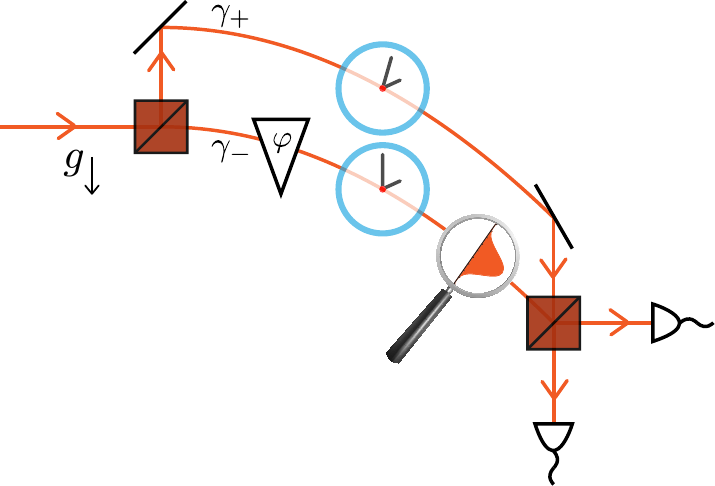}
    \caption{An atomic clock, represented with a Gaussian wave packet with internal degrees of freedom, is in a superposition of two freely falling trajectories with gravitational acceleration $g$, and subsequently it is recombined and measured. The setup is made of two beam splitters, a phase shifter of controllable phase $\varphi$, and two detectors. The upper trajectory is called $\gamma_+$ and the lower $\gamma_-$.}
    \label{fig:FF}
\end{figure}

We now study the dynamics of the quantum clock when it traverses a freely falling interferometer. The interferometer is made of two beam splitters, a phase shifter of controllable phase $\varphi$, and two detectors, as depicted in Figure~\ref{fig:FF}. The atom is in a superposition of two trajectories $\gamma_\pm$, each composed of a vertical path and a parabolic path. This setup provides a way to measure combined effects of quantum theory and general relativity, since it realises a superposition of different gravitational time dilations: the clock in each path of the interferometer measures a different proper time, and this implies a difference in the phase accumulated in each trajectory, which affects the interferometric probabilities, thus being a measurable effect. \cedit{For the interested reader, Appendix \ref{sec:app_floor} is dedicated to the study of this setup in the presence of a floor, modelled as an infinite barrier of potential.}

We assume that the acceleration and velocity are the same in each vertical path, so that their contribution cancels, and we focus on the parabolic paths only.

The Hamiltonian for each parabolic path is
\begin{align}
\label{eqApp:q_hamiltonian_balanced}
    \hat H = mc^2 & + \frac{\hat p^2}{2m} + mV_{F}(\hat x) \notag \\ & + \hat H_{int}\left(1+\frac{V_{F}(\hat x)}{c^2} -\frac{\hat p^2}{2m^2c^2}\right)\,,
\end{align}
where we take a linear approximation of the gravitational potential, namely
\begin{equation} \label{eq:FF_potential}
    V_{F}(\hat x) = g (\hat x - x_0) + V_N(x_0),
\end{equation}
where $V_N(x)$ is the Newtonian potential.

\fedit{The momentum operator $\hat p$ in the Hamiltonian of Eq.~\eqref{eqApp:q_hamiltonian_balanced} is only the vertical component of the momentum: we neglect the horizontal dynamics of the quantum state, because it does not influence the proper time measured by the clock}. %Therefore, we discard the free atom horizontal evolution.

In order to describe the external degrees of freedom of the atom, we employ Gaussian wave packets centered at the two heights of the interferometer $x_\pm$ with an initial \fedit{variance in position basis} $\sigma$. The explicit expression is
\begin{align} \label{eq:gaussian_state}
    \ket{\psi_\pm} &= \frac{1}{(2 \pi \sigma^2)^{1/4}}\int dx \ e^{-\frac{(x-x_\pm)^2}{4\sigma^2}} \ket{x} \nonumber \\
    &= \left(\frac{2\sigma^2}{\pi\hbar^2}\right)^{1/4}\int dp \ e^{-\frac{p^2 \sigma^2}{\hbar^2} - \frac{i p x_\pm}{\hbar}} \ket{p}.
\end{align}
The initial state inside the interferometer, after the first beam splitter and the phase shifter, is
\begin{equation}\label{eq:initial_state_MZ}
    \ket{\psi_0} = \frac{\ket{\psi_+} + e^{i\varphi}\ket{\psi_-}}{\sqrt{2}}\ket{\tau_{in}},
\end{equation}
where $\ket{\psi_\pm}$ are the Gaussian states of Eq.~\eqref{eq:gaussian_state} and $\ket{\tau_{in}}$ is the initial internal state of the clock, which we choose to be
\begin{equation}\label{eq:initial_internal_state}
    \ket{\tau_{in}} = \frac{\ket{0}+\ket{1}}{\sqrt{2}}.
\end{equation}

The state inside the interferometer after a time $\Delta t$ is 
\begin{align}\label{eq:MZ_evolved_state}
    \ket{\psi} & = U\ket{\psi_0} \\  
    & = \frac{\ket{+_0}\ket{0} + \ket{+_1} \ket{1} + e^{i\varphi}(\ket{-_0}\ket{0} + \ket{-_1}\ket{1})}{2}\,. \notag
\end{align}
where the states $\ket{\pm_i}$ are Gaussian states \fedit{characterised by the following \cedit{phase}, mean momentum, mean position, and variance in position basis} after the evolution of a time $\Delta t$ generated by the Hamiltonian in Eq.~\eqref{eqApp:q_hamiltonian_balanced}:
\begin{align}
    \phi_{\pm,i} (x) =&  -\frac{\Delta t E_i}{\hbar} - \frac{\Delta t\, m\, V_{F}(x)}{\hbar}\left(1+\frac{E_i}{mc^2}\right) \notag \\ &-\frac{mg\Delta t^3}{3\hbar}\left[1+\frac{E_i}{mc^2}- \left(\frac{E_i}{mc^2}\right)^2\right], \label{eq:FF_complete_phase}\\
    \braket{p}_{\pm,i} &= m g \Delta t \left(1 + \frac{E_i}{mc^2}\right), \label{eq:FF_complete_velocity}\\
    \braket{x}_{\pm,i} &= x_\pm-\frac{g \Delta t^2}{2}\left[1- \left(\frac{E_i}{mc^2}\right)^2\right], \label{eq:FF_complete_position}\\
    \Sigma^2_i &= \sigma^2 + \left[\frac{\hbar \Delta t}{2m\sigma}\left(1 - \frac{E_i}{mc^2}\right)\right]^2. \label{eq:FF_complete_variance}
\end{align}
We can now quantify how much information on the gravitational acceleration is contained in the state at the end of the interferometer, through the use of the Quantum Fisher Information (QFI---see Appendix \ref{sec:app_QFI} for details).
The QFI for $g$ of this state in the limit of long times and small internal energies is 
\begin{equation} \label{eq:FF_QFI_asymp}
    G^{F}_{\text{asy}}(g) \overset{\Delta t\to \infty}\simeq \, \frac{g^2 \Delta E^2}{36\, \hbar ^2c^4 }\, \Delta t^6.
\end{equation}
This scaling in $\Delta t^6$ is a consequence of gravitational time dilation, since it comes from the coupling between the gravitational potential and the internal energies. \rev{If there was an internal clock that was not coupled to the gravitational field, namely if we did not consider gravitational time dilation, this scaling would be absent: the phase term that scales as $\Delta t^3$ in Eq.~\eqref{eq:FF_complete_phase}, which is responsible for the $\Delta t^6$ scaling, would be only a global phase, and would thus be unobservable. Without time dilation, the scaling of the QFI would be $\Delta t^4$, as shown in Ref.~\cite{Kritsotakis2018}.} Therefore, gravitational time dilation enhances the sensitivity of the setup for long interferometric times, introducing a time scaling that would be otherwise absent.

\begin{figure}[h!]
    \includegraphics[width=0.95\columnwidth]{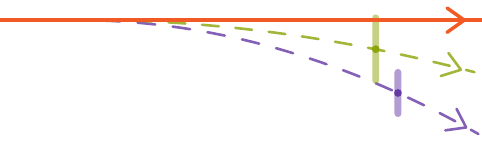}
    \caption{Trajectories of the states $\ket{+}_0$ (green line) and $\ket{+}_1$ (violet line), which are centered in the same interferometric branch, but with different internal energies. The paths differ in mean position, mean momentum and variance, according to Eqs.~(\ref{eq:FF_complete_velocity}~-~\ref{eq:FF_complete_variance}).}
    \label{fig:FF_approximations}
\end{figure}

From now on, we restrict our calculations in a specific regime. In particular, we require that the spread of the wave packet is smaller than the vertical distance between the trajectories (i.e. there is no overlap between the quantum states of the atom in the two trajectories):
\begin{equation}\label{eq:MZ_Sigma_condition}
    \Sigma \ll h = x_+ - x_-,
\end{equation}
where $\Sigma$ is the \fedit{width of the wave function of the atom in position basis}. Moreover, two states \fedit{following the same path, but} with different internal energies, such as $\ket{+}_0$ and $\ket{+}_1$, have slightly different trajectories: there is a difference in the mean positions, mean velocities and variances, as it is seen in Eqs.~(\ref{eq:FF_complete_velocity}-\ref{eq:FF_complete_variance}), and depicted in Figure~\ref{fig:FF_approximations}.

We study the regime where these effects are negligible, namely where there is a single trajectory for both internal energy states. This allows us to perform an interferometric measurement without having to account for the different trajectories of each internal degree of freedom. \rev{Moreover, we keep only the leading term in the internal energies.} Therefore, we approximate \cedit{phase}, mean momentum, mean position and variance with
\begin{align}
    \phi_{\pm,i} (x)=& -\frac{\Delta t E_i}{\hbar} - \frac{\Delta t mV_{F}(x)}{\hbar}\left(1+\frac{E_i}{mc^2}\right) \notag \\ &-\frac{mg\Delta t^3}{3\hbar}\left(1+\frac{E_i}{mc^2}\right), \label{eq:FF_phase}\\
    \braket{p}_{\pm} &= m g \Delta t, \label{eq:FF_velocity}\\
    \braket{x}_{\pm} &= x_\pm-\frac{g \Delta t^2}{2}, \label{eq:FF_position}\\
    \Sigma^2 &= \sigma^2 + \left(\frac{\hbar \Delta t}{2m\sigma}\right)^2. \label{eq:FF_variance}
\end{align}

If the condition \fedit{on the localization of the wave function of Eq.~\eqref{eq:MZ_Sigma_condition} and the previous conditions} are satisfied, the state in Eq.~\eqref{eq:MZ_evolved_state} represents an atom in a superposition of two branches, where in each branch, the atom is centered in a trajectory, with a negligible \fedit{overlap with the quantum state centered in} the other trajectory, and the evolution is the same regardless of the internal state of the clock. \fedit{In Appendix~\ref{sec:app_approx} we show that such regime exists, and we explicitly give a range of the parameters satisfying all the conditions.}

In these approximations, the QFI reads
\begin{align} \label{eq:FF_QFI}
    G^{F}(g) &= \frac{\left(1+z_0\right)^2+\left(1+z_1\right)^2}{2} \left(\frac{m \Delta t}{\hbar }\right)^2 \left(4 \Sigma ^2+h^2\right)\nonumber \\ 
    &+\left(\frac{m \Delta t \Delta z}{\hbar}\right)^2 \left(\frac{g \Delta t^2}{6}+\bar h\right)^2-\left(\bar z+\frac{3}{4}\right)\frac{\Delta t^4}{\sigma ^2},
\end{align}
where $z_i = \frac{E_i}{mc^2}$, $\Delta z = z_1 - z_0$, $\bar z = \frac{z_0 + z_1}{2}$, $h = x_+ - x_-$ and $\bar h = \frac{x_+ + x_-}{2} - x_0$.

We notice that the presence of gravitational time dilation not only enhances the time scaling for long interferometric times, but also increases the coefficients of the other powers of $\Delta t$.

Since most of the interferometric measurements regard the external degrees of freedom only, namely they are obtained through interference of the two interferometric paths and subsequent measurement of each path, we study what information is contained in the reduced state, where the internal degrees of freedom are ignored.

The reduced state is
\begin{equation}
 \rho_{red} = \frac{1}{2}(\ket{\phi_0}\bra{\phi_0} + \ket{\phi_1}\bra{\phi_1}),
\end{equation}
where $\ket{\phi_i} = \frac{1}{\sqrt{2}}(\ket{+_i} + e^{i\varphi}\ket{-_i}$.

We need to diagonalize the reduced state in order to calculate its QFI. The Hilbert space of this state is infinite-dimensional, since the position can assume infinite values, meaning that we would have to diagonalize an infinite-dimensional matrix. For simplicity, we approximate the Hilbert space to be 2-dimensional, and the state to be a qubit. In order to achieve this, we approximate the states $\ket{\pm_i}$ with $\ket{\pm_i} = e^{i\phi_{\pm,i}} \ket{\pm}$, where $\phi_{\pm i}$ are the phases of the original states evaluated in the center of the trajectory, namely $\phi_{\pm,i} := \phi_{\pm,i}\left(x_\pm - \frac{g\Delta t^2}{2}\right)$.

With this approximation, we are discarding every effect related to the spread of the wave packet, meaning that we ignore the phases near the center of the trajectory, and the variance of the wave packet $\Sigma$, and thus this operation can be regarded as a semiclassical approximation.

We obtain 
\begin{align}\label{eq:FF_QFI_reduced}
    G^{F}_{red}(g) &= \left(\frac{m \Delta z \Delta t h}{2\hbar}\right)^2 +  \\  &+\left(\frac{m\Delta t h}{\hbar}\left(1+\bar z\right)\right)^2 \cos^2{\left(\frac{m \Delta z \Delta V_{F} \Delta t}{2\hbar} \right)}. \nonumber
\end{align}
The factors of the total QFI in Eq.~\eqref{eq:FF_QFI} that scale as $\Delta t^4$ and $\Delta t^6$ disappear. One of the $\Delta t^4$ factor of Eq.~\eqref{eq:FF_QFI} disappears because of the semiclassical approximation, since we are ignoring every contribution coming from the spread of the wave packet, such as the term that depends on $\sigma$. The other $\Delta t^4$ and $\Delta t^6$ terms, nevertheless, are not related to the spread of the wave packet, meaning that their disappearance is not related to the approximation: they disappear because internal degrees of freedom have been ignored.

This suggests that measuring only the external degrees of freedom does not exploit all the information contained in the state.

A simple interferometric measurement that can be performed is 
\begin{equation} \label{eq:FF_measurement}
    \ket{D_\pm} = \frac{\ket{+} \pm \ket{-}}{\sqrt{2}}, \qquad \ket{\pm} = \ket{\pm_i}\Big|_{E_i=0},
\end{equation}
namely the two paths are interfered and each output is measured (see Figure \ref{fig:FF}). The internal degrees of freedom are not measured, but the effects of the internal clock is visible in the phase acquired by the state. This measurement is sensitive to the phase difference between the branches, while discarding any effect that does not depend on the time dilation, since it projects on states where any other phase is already included, and thus the other phases do not influence the measurement probabilities. \rev{The probabilities are \rev{calculated} projecting the state Eq.~\eqref{eq:MZ_evolved_state} on the states of Eq.~\eqref{eq:FF_measurement}, after applying the approximations described above. The result is }
\begin{equation} \label{eq:prob_one_particle_gaussian}
    P_\pm \sim \frac{1}{2} \left( 1 \pm \cos{\left(\frac{\Delta E \Delta V_{F} \Delta t}{2\hbar c^2}\right)}\cos{\left(\frac{\bar E \Delta V_{F} \Delta t}{\hbar c^2} + \varphi \right)}\right).
\end{equation}
As explained in Appendix \ref{sec:app_QFI}, one can estimate the information on $g$ that can be extracted from these probabilities, using the Fisher Information (FI). The FI is
\begin{equation}\label{eq:FI_FF}
    F(g) = \left(\frac{\Delta E h \Delta t}{2 \hbar c^2}\right)^2. \\
\end{equation}

It does not show the $\Delta t^6$ scaling, as a consequence of ignoring the internal degrees of freedom, but also the $\Delta t^4$ does not appear. \cedit{The reason is that we defined the measurement to be sensitive to $g$ only as a gravitational time dilation effect}, while the reduced state has an additional dependence on $g$ that does not depend on the coupling with the internal energies. Therefore, our measurement is ignoring that information, and this loss of information is quantified by the difference between the QFI of the reduced state and the FI of the measurement.

\section{Mach-Zehnder interferometer}
\label{s:mzi}
\begin{figure}[ht]
    \centering
    \includegraphics[width=\columnwidth]{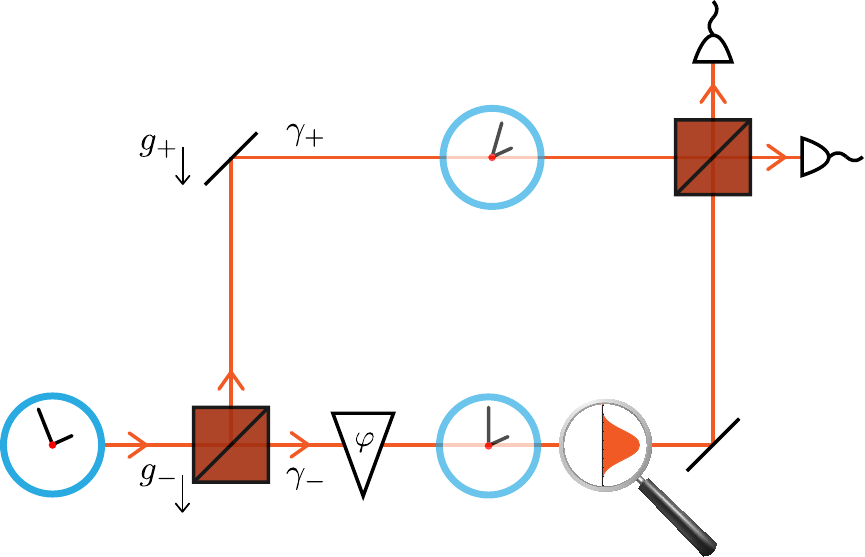}
    \caption{An atomic clock, represented with a Gaussian wave packet with internal degrees of freedom, in a Mach-Zehnder interferometer vertically placed in a piecewise linear potential, with accelerations $g_\pm$. The setup is made of two beam splitters, a phase shifter of controllable phase $\varphi$, and two detectors. The upper trajectory is called $\gamma_+$ and the lower $\gamma_-$.}
    \label{fig:Mach-Zehnder}
\end{figure}
In this section we analyze the scenario where the atom evolves in a Mach-Zehnder interferometer, as depicted in Figure~\ref{fig:Mach-Zehnder}. To keep the atom at a constant height, we introduce a potential that balances the gravitational attraction. \rev{For example, this potential could be an atomic trap, that keeps the atom still but does not couple to the internal clock via time dilation.} Therefore, the Hamiltonian for each horizontal path is
\begin{equation}\label{eq:q_hamiltonian_balanced}
    \hat H = mc^2 + \frac{\hat p^2}{2m} + \hat H_{int}\left(1+\frac{V_N(\hat x)}{c^2} -\frac{\hat p^2}{2m^2c^2}\right).
\end{equation}
%Without the internal clock, namely if $\hat H_ {int}=0$, the atom would be free, meaning that any dynamics different from the free evolution is caused by the internal degrees of freedom. 
\fedit{Notice that the gravitational field acts exclusively on the internal degrees of freedom via} gravitational time dilation.

\begin{figure}[t]
     \centering
     \begin{subfigure}{0.49\columnwidth}
         \centering
         \includegraphics[width=\textwidth]{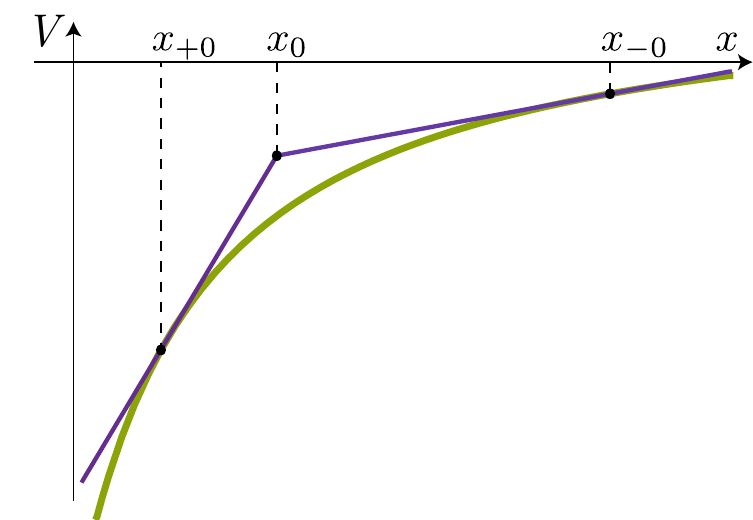}
         \caption{}
         \label{fig:potential}
     \end{subfigure}
     \hfill
     \begin{subfigure}{0.49\columnwidth}
         \centering
         \includegraphics[width=\textwidth]{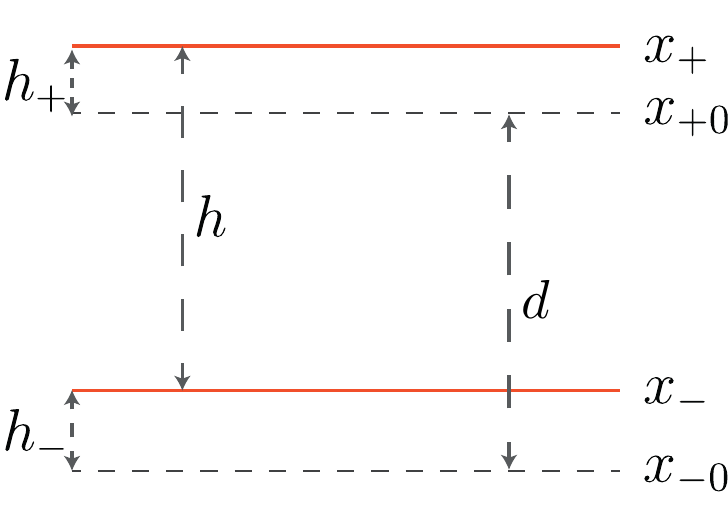}
         \caption{}
         \label{fig:potential_situation}
     \end{subfigure}
     \caption{On the left, the piecewise linear approximation (violet line) of the gravitational potential Eq.~\eqref{eq:MZ_cl_potential} (green line). On the right, the representation of the quantities described in Eqs.~(\ref{eq:potential_condition},~\ref{eq:H_condition}).}
\end{figure}

In order to solve the dynamics, we may refine the previous approximation of a linear potential, using a piecewise linear potential: \fedit{this means that in each horizontal trajectory the dynamics is linear}. We expand the Newtonian potential with a Taylor series centered in two points $x_{\pm 0}$, thus obtaining
\begin{singlespace}
\begin{equation}\label{eq:MZ_cl_potential} V_{MZ}(x) = 
\begin{cases} 
    g_+ (x-x_{+0}) + V_N(x_{+0}) \quad \text{if $x>x_0$}, \\
    g_- (x-x_{-0}) + V_N(x_{-0}) \quad \text{if $x<x_0$}, \\
\end{cases}
\end{equation}
\end{singlespace}\noindent
where $V_N(x)$ is the Newtonian potential, $x$ is the distance from the center of the Earth, $x_0$ is the $x$-coordinate of the intersection of the two lines, and $g_\pm = V_N^\prime(x_{\pm0})$ are the gravitational accelerations at $x_{\pm0}$. The potential is represented in Fig.~\ref{fig:potential}.

This piecewise linear approximation holds for heights that are small with respect to the scale over which the potential changes, namely
\begin{equation}\label{eq:potential_condition}
    h_\pm \equiv x_\pm - x_{\pm0} \ll d \equiv x_{+0} - x_{-0}.
\end{equation}
Using this condition, we obtain that the distance $h$ between the interferometer trajectories is
\begin{equation}\label{eq:H_condition}
    h = x_+ - x_- \sim d.
\end{equation}
This situation is represented in Figure~\ref{fig:potential_situation}.

To obtain a potential \fedit{acting on the quantized path degree of freedom of the atom}, we quantize the position operator in each linear piece, 
%but we do not quantize the Heaviside step functions that divide the linear pieces: this corresponds to the approximation where an 
\fedit{and we assume that the atom in each} trajectory is localized enough for its spread to be small compared to the distance between the two trajectories, as in Eq.~\eqref{eq:MZ_Sigma_condition}.

\fedit{The resulting potential is}
\begin{align}\label{eq:q_potential}
    V_{MZ}(\hat x) = &\theta \left(x-x_0\right) \left(g_+ (\hat x-x_{+0}) + V_N(x_{+0})\right) + \nonumber\\ &\theta \left(x_0-x\right) \left(g_- (\hat x-x_{-0}) + V_N(x_{-0})\right).
\end{align}

Consequently, the derivative of the potential with respect to the position operator $\hat x$ is
\begin{equation}
    \dot {\hat V}_{MZ}(x) \equiv \frac{d}{d \hat x}V_{MZ}(\hat x) = \theta(x - x_0) g_+ + \theta(x_0 - x) g_-,
\end{equation}
where the derivative does not affect the Heaviside step functions. %since they are not operators.

Let’s consider the initial state to be Eq.~\eqref{eq:initial_state_MZ}, in the same hypothesis as above, namely we consider a superposition of the atom centered in two trajectories, with null probability to be found in the other trajectory, and the evolution does not change between different internal degrees of freedom. The evolved state after a time $\Delta t$ is formally
analogous to Eq.~\eqref{eq:MZ_evolved_state}, where $\ket{\pm_i}$ are Gaussian states with

\begin{align}
    \phi_{\pm,i} (x)&= -\frac{\Delta t E_i}{\hbar} - \frac{\Delta t mV_{MZ}(x)}{\hbar}\frac{E_i}{mc^2}, \\
    \braket{p} &= 0 \label{eq:MZ_velocity}\\ 
    \braket{x}_{\pm} &= x_\pm, \label{eq:MZ_mean_value}\\
    \Sigma^2 &= \sigma^2 + \left(\frac{\hbar \Delta t}{2 m \sigma}\right)^2. \label{eq:MZ_variance}
\end{align}

We estimate the sensitivity of this setup to the difference in gravitational acceleration $\Delta g = g_- - g_+$ and to the mean gravitational acceleration $\bar g = \frac{g_+ + g_-}{2}$, by calculating the Quantum Fisher Information for these two parameters. Under the approximations described above, the QFIs read
\begin{align}
    &G^{MZ}(\Delta g) = \left(\frac{\Delta t}{4\hbar c^2}\right)^2 \left((E_0^2+E_1^2)(8(\Sigma^2 + \bar h^2) + \Delta E^2 \Delta h^2\right), \label{eq:QFI_MZ_Deltag}\\
    &G^{MZ}(\bar g) = \left(\frac{\Delta t}{\sqrt{2}\hbar c^2}\right)^2 \left((E_0^2+E_1^2)(4\Sigma^2 + \Delta h^2) + 2\Delta E^2 \bar h^2\right), \label{eq:QFI_MZ_g}
\end{align}
where $\bar h = \frac{h_+ + h_-}{2}$ and $\Delta h = h_+ - h_-$. As we anticipated, the dependence on $\Delta g$ and $\bar g$ is present due to a gravitational time dilation effect, since without the internal energies, namely $E_i=0$, the QFI would be null.

As in the previous case, we study the QFI of the reduced state, to see whether ignoring the internal degrees of freedom leads to an important loss of information. We obtain

\begin{align} \label{eq:QFI_MZ_Deltag_reduced}
    G^{MZ}_{red}(\Delta g) &= \left(\frac{\Delta E \bar h \Delta t}{2 \hbar c^2}\right)^2 + \nonumber \\ &+\left(\frac{\bar E \bar h \Delta t}{\hbar c^2}\right)^2 \cos^2{\left(\frac{\Delta E \Delta V_{MZ} \Delta t}{2\hbar c^2} \right)}, \\
    G^{MZ}_{red}(\bar g) &= \left(\frac{\Delta E \Delta h \Delta t}{2 \hbar c^2}\right)^2 + \nonumber \\ &+ \left(\frac{\bar E \Delta h \Delta t}{\hbar c^2}\right)^2 \cos^2{\left(\frac{\Delta E \Delta V_{MZ} \Delta t}{2\hbar c^2} \right)}. \label{eq:QFI_MZ_reduced_g}
\end{align}
Hence, ignoring the external degrees of freedom does not erase the information about $\bar g$ and $\Delta g$ that is contained in the state: the time scaling is preserved.

Consequently, the Fisher Information (FI) for $\Delta g$ and $\bar g$ is
\begin{align}
    F^{MZ}(\Delta g) &= \left(\frac{\Delta E \bar h \Delta t}{2 \hbar c^2}\right)^2, \label{eq:FI_MZ_Deltag}\\
    F^{MZ}(\bar g) &= \left(\frac{\Delta E \Delta h \Delta t}{2\hbar c^2}\right)^2.\label{eq:FI_MZ_Barg}
\end{align}
\section{Conclusion}
\label{s:out}
\medit{We have performed a  metrological analysis of two atomic clock interferometers: a \cedit{freely falling interferometer and a Mach-Zehnder interferometer}.}
\medit{Our results show that gravitational time dilation can be used as a resource to improve precision, \cedit{and this effect becomes significant} in the limit of long interferometric times.} 

\medit{In particular, we have shown that gravitational time dilation enhances the time scaling of the information that a freely falling interferometer can extract about the gravitational acceleration, going from a $\Delta t^4$ scaling to a $\Delta t^6$ one. To achieve this performance, i.e. to extract the whole information contained in the setup,  a joint measure of internal and external degrees of freedom should be performed. If we  discard the internal degrees of freedom, the $\Delta t^6$ scaling is lost.}

\medit{We have also shown that gravitational time dilation makes Mach-Zehnder interferometer sensitive to the difference in the gravitational acceleration between the locations of the two arms, as well as to their mean value. We have quantified the information contained in the interferometric state, and found an optimal measurement  
able to extract the maximum achievable information.} 

Finally, we have shown (see Appendix \ref{sec:app_floor}) that the gravitational time dilation does not enhance the time scaling for the freely falling interferometer when the floor is considered. \fedit{This means that, in order to benefit from the enhancement provided by gravitational time dilation, the measurements in the free-fall interferometer should be performed far from the potential barrier corresponding to the floor.}

\medit{The practical relevance of our findings is still limited, given the currently available coherence of quantum clocks~\cite{rv22}. On the other hand, our results clearly show that gravitational time dilation may indeed represent a resource for quantum metrology, and provide a novel fundamental avenue to test the interface between quantum theory
and gravity.}

\section*{Code availability}
The \emph{Mathematica} notebook is included in the arXiv repository of this paper.
\acknowledgements{C.C. was supported by the Austrian Science Fund (FWF) through BeyondC (F7103-N48), the John Templeton Foundation (ID\# 61466) as part of The Quantum Information Structure of Spacetime (QISS) Project (qiss.fr) and the European Commission via Testing the Large-Scale Limit of Quantum Mechanics (TEQ) (No. 766900) project. F.G. acknowledges support from Perimeter Institute for Theoretical Physics. Research at Perimeter Institute is supported in part by the Government of Canada through the Department of Innovation, Science and Economic Development and by the Province of Ontario through the Ministry of Colleges and Universities. \medit{M.G.A.P. is member of INdAM-GNFM}.}

\appendix
\section{A few words on estimation theory}\label{sec:app_QFI}
Here, we briefly review some tools of classical and quantum estimation theory, following Refs.~\cite{qm2,liu,rem}. Classical estimation theory addresses the problem of finding the best estimator to process the data obtained with a given measurement. 
Given a parameter of interest $\lambda$ and a sample space $\chi = \{x\}$, namely a set of all possible outcomes of an experiment, an estimator $\hat \lambda$ is a function from the sample space $\chi$ to the space of the possible values of $\lambda$. The variance $\Delta^2(\lambda)$ of every unbiased estimator is subjected to the Cramér-Rao bound
\begin{equation}\label{eq:cl_cramer_rao}
    \Delta^2(\lambda) \geq \frac{1}{M F(\lambda)},
\end{equation}
where $M$ is the number of measurements performed and $F(\lambda)$ is the Fisher Information (FI) of the estimator. For a discrete set of outcomes $\chi$, the Fisher Information is defined as
\begin{equation}\label{eq:FI}
    F(\lambda) = \sum_{x \in \chi} \frac{(\partial_\lambda P_\lambda(x))^2}{P_\lambda (x)},
\end{equation}
where $P_\lambda(x)$ is the probability of obtaining a certain outcome $x$ for a fixed value of the parameter of interest $\lambda$.
The Fisher Information quantifies the information that the measured quantity $X$, associated to the sample space $\chi$, carries about the parameter $\lambda$: the bigger the Fisher Information for a certain parameter is, the lesser its variance is. This is a way to determine the best estimator for a given set of data, namely the one that saturates the Cramér-Rao inequality.

The same concepts may be carried over to quantum theory. One can find a quantum Cram{\'e}r-Rao bound
\begin{equation} \label{eq:q_cramer_rao}
    \Delta^2(\lambda) \geq \frac{1}{M G(\lambda)},
\end{equation}
where $G(\lambda)$ is the Quantum Fisher Information (QFI). Saturating this bound provides the best estimator, as in the classical setting, but it also defines what the best measurement to be performed is. An intuitive explanation is that, contrary to classical theory, in quantum mechanics one choice of measurement can be better than another due to the non-commutativity of observables: there is a bound on the information that can be extracted about an eigenstate of an observable $\hat A$, from a measurement of an observable $\hat B$ that does not commute with $\hat A$.

The QFI for a pure state can be shown to be
\begin{equation}\label{eq:QFI_pure}
    G(\lambda) = 4\left(\braket{\partial_\lambda \psi | \partial_\lambda \psi} - |\!\braket{\partial_\lambda \psi| \psi}\!|^2\right),
\end{equation}
where $\ket{\partial_\lambda \psi} = \sum_k \partial_\lambda \psi_k \ket{k}$ with $\{\ket{k}\}$ a basis of the Hilbert space that does not depend on the parameter $\lambda$.

Similarly, for a mixed state $\rho = \sum_i \rho_i \ket{\psi_i}\bra{\psi_i}$, we have
\begin{align}\label{eq:QFI_mixed}
    G(\lambda) =& \sum_{\rho_i} \frac{(\partial_\lambda \rho_i)^2}{\rho_i} + 4\sum_{\rho_i} \rho_i \braket{\partial_\lambda \psi_i| \partial_\lambda \psi_i} \nonumber\\ &- 8 \sum_{\rho_i,\rho_j} \frac{\rho_i \rho_j}{\rho_i + \rho_j} |\! \braket{\partial_\lambda \psi_i | \psi_j} \! |^2.
\end{align}

\rev{In the main text, Eq.~\eqref{eq:FI} has been explicitely used to calculate the FI of the measurement probabilities for the freely falling interferometer (Eq.~\eqref{eq:prob_one_particle_gaussian}), obtaining Eq.~\eqref{eq:FI_FF}. The probabilities for the Mach-Zehnder interferometer have the same form of Eq.~\eqref{eq:prob_one_particle_gaussian}, but with a different potential. The result is that the FI for $\bar g$ and $\Delta g$ are slightly different, and they are given in Eqs.~(\ref{eq:FI_MZ_Deltag},~\ref{eq:FI_MZ_Barg}). Similarly, Eq.~\eqref{eq:QFI_pure} has been used explicitly to calculate Eqs.~(\ref{eq:FF_QFI_asymp},~\ref{eq:FF_QFI},~\ref{eq:FF_QFI_reduced},~\ref{eq:QFI_MZ_Deltag}~-~\ref{eq:QFI_MZ_reduced_g}) with the aid of \emph{Mathematica} software.}
\section{The approximation regime}\label{sec:app_approx}
Here we provide few more details on the regime that is considered in the analysis of the freely falling interferometer and of the Mach-Zehnder interferometer in the main text.
The localization condition of Eq.~\eqref{eq:MZ_Sigma_condition} is equivalent to the assumptions
\begin{equation}
    \sigma \ll h, \qquad \sigma \gg \frac{\hbar}{m} \frac{\Delta t}{h},
\end{equation}
while the approximations employed to write  Eqs.~(\ref{eq:FF_phase}-\ref{eq:FF_variance}) are equivalent to
\begin{align}
    m g \Delta t \frac{E_i}{mc^2} \ll \Sigma_p \equiv \frac{\hbar}{\Sigma}, \label{eq:appa_velocity_condition}\\
    \frac{g\Delta t^2}{2}\frac{E_i}{mc^2}\ll \Sigma, \label{eq:appa_position_condition}\\
    \frac{\hbar \Delta t}{m\sigma} \sqrt{\frac{E_i}{mc^2}} \ll \Sigma \label{eq:appa_variance_condition_1} .
\end{align}
In the hypothesis of employing ${}^{88}$Sr atoms, which is an atomic species whose electronic transitions are currently employed in atomic clocks, approximated values for the parameters are $m_{Sr}\sim10^{-25}$~kg, $\Delta E \sim 2.8$~eV. Thus, one finds that a possible set of parameters that satisfies the above inequalities is $\Delta t \sim 10$~s, $h\sim 1$~cm, $\sigma \sim 10^{-4}$~m$^2$ or for longer times $\Delta t \sim 100$~s, $h\sim 1$~cm, $\sigma \sim 10^{-3}$~m$^2$.
\section{Freely falling interferometers in the presence of a floor}\label{sec:app_floor}
\label{s:ffif}
In this appendix, \fedit{we consider} an atomic clock in a superposition of two freely falling trajectories, like in Section~\ref{s:ffi}, \fedit{but, in addition to the gravitational potential, we also have an infinite barrier of potential modelling the presence of the floor next to the detectors at the end of the interferometer.} 

\fedit{This situation is also referred to as} a \emph{quantum bouncer}~\cite{sakurai}. Here, we generalize the \fedit{standard} treatment to atoms with internal degrees of freedom. \fedit{We follow} the analysis of Ref.~\cite{seveso}.

We \fedit{consider} a Hamiltonian
\begin{equation}
    \hat H = \frac{\hat p^2}{2m} + m V_f(\hat x) + \hat H_{int}\left(1 + \frac{\hat V_{F}(\hat x)}{c^2}\right),
\end{equation}
\fedit{where, for simplicity, we have discarded the rest mass term and ignored special relativistic effects}.

Here, $V_{F}(\hat x)$ is the linear gravitational potential of Eq.~\eqref{eq:FF_potential}, and $V_f(\hat x)$ is the potential modified because of the floor, \fedit{namely
\begin{singlespace}
\begin{equation}
    V_f(x) = 
\begin{cases} 
    V_{F}(x) &\text{if $x>0$} \\
    \infty  &\text{if $x=0$}. \\
\end{cases}
\end{equation}
\end{singlespace}}

\fedit{The presence of the floor does not influence the general relativistic time dilation on the clock Hamiltonian $\hat H_{int}$, but imposes boundary conditions on the path degrees of freedom: the eigenvalues of the Hamiltonian $E_{i,n}$, corresponding to the eigenstates $\ket{\Psi_{i,n}} = \ket{\psi_{i,n}}\ket{E_i}$, are discrete}, as calculated explicitly in the following. 

The eigenvalue equation is
\begin{equation}
    \hat H \ket{\psi}\ket{E_i} = E \ket{\psi} \ket{E_i}.
\end{equation}
If we let the internal degrees of freedom act on the internal Hamiltonian, and project on $\ket{x}\ket{E_i}$, we obtain the time-independent Schrödinger equation
\begin{align}
    &\left( -\frac{\hbar^2}{2m} \partial^2 + m\left(g\left(x-x_0\right) + V(x_0)\right)\left(1+\frac{E_i}{mc^2}\right) + E_i\right) \psi(x) = \nonumber \\ &= E \ \psi(x).
\end{align}
Then, we perform the substitutions 
\begin{align}
    y &= k_i\left(x-a_i\right), \\
    a_i &= \frac{E-E_i}{mg\left(1+\frac{E_i}{mc^2}\right)} + x_0 - \frac{V(x_0)}{g}, \\
    k_i &= \frac{1}{l_i} = \left(\frac{2 m^2 g}{\hbar^2}\left(1+\frac{E_i}{mc^2}\right)\right)^{\frac{1}{3}}, \label{eq:FFwF_gravitational_l}\\
    \psi(x) &= \sqrt{k_i} \ \phi(y).
\end{align}
After the substitutions, the Schrödinger equation is written as an Airy equation, namely
\begin{equation}
    \phi(y)^{\prime\prime} = y \ \phi(y),
\end{equation}
whose solutions are linear combinations of the Airy functions $\text{Ai}(y)$ and $\text{Bi}(y)$. We select the ones that are finite for $y \to \infty$, namely
\begin{equation}
    \phi(y) = A \ \text{Ai}(y), 
\end{equation}
where
\begin{equation}
    \text{Ai}(y) = \frac{1}{2\pi} \int_{-\infty}^{+\infty} du \ e^{i(yu + u^3/3)}.
\end{equation}
Therefore, the wave functions are
\begin{equation}
    \psi(x) = N \ \text{Ai}(k_i(x-a_i)).
\end{equation}

The boundary condition $\psi(0) = 0$, stemming from the presence of the floor, selects only discrete values of $a_i$, namely
\begin{equation}
    a_{i,n} = - l_i z_n,
\end{equation}
where $z_n<0$ is the $n$-th zero of the Airy function. 

This results in discrete eigenvalues 
\begin{equation}
    E = E_{i,n} = m (-g (z_n l_i + x_0) + V(x_0))\left(1+\frac{E_i}{mc^2}\right) + E_i,
\end{equation}
and discrete eigenstates
\begin{equation}\label{eq:FFwF_energy_eigenvalues}
    \ket{\Psi_{i,n}} = \ket{\psi_{i,n}}\ket{E_i},
\end{equation}
with eigenfunctions
\begin{equation}
    \psi_{i,n}(x) = N_{i,n} \ \text{Ai}\left(\frac{x}{l_i} + z_n\right), \qquad N_{i,n} = \frac{1}{\sqrt{l_i}\text{Ai}^\prime(z_n)}.
\end{equation}

The projection of the initial state of Eq.~\eqref{eq:initial_state_MZ} on the eigenstates of the Hamiltonian is
\begin{align}
    \ket{\psi_0} & = \frac{\ket{\psi_+} + e^{i\varphi}\ket{\psi_-}}{\sqrt{2}}\ket{\tau_{in}} \notag \\ & = \sum_{i,n} c_{i,n} \ket{\psi_{i,n}}\ket{E_i},
\end{align}
where the coefficients can be derived in the same way as Ref.~\cite{seveso}. The result is
\begin{align}
    c_{i,n} &= \frac{\braket{\psi_{i,n} | \psi_+} +e^{i\varphi} \braket{\psi_{i,n} | \psi_-}}{\sqrt{2}} \braket {E_i | \tau_{in}} \notag \\ &= \frac{c_{i,n}^+ +e^{i\varphi} c_{i,n}^-}{2}, \label{eq:FFwF_coefficient}
\end{align}
with
\begin{align}
c_{i,n}^\pm &= \left(\frac{1}{2\pi\sigma^2}\right)^{\frac{1}{4}} N_{i,n} \exp{\left(\frac{l_{\sigma,i}^2}{2}(z_n + l_{\pm,i})+\frac{l_{\sigma,i}^2}{12}\right)} \times \nonumber \\ &\times \text{Ai}\left(z_n + l_{\pm,i} + \frac{l_{\sigma,i}^2}{4}\right),
\end{align}
where $l_{\pm,i} = \frac{l_\pm}{l_i}$ and $l_{\sigma,i} = \frac{\sigma}{l_i}$.

If we further assume that the initial spread of the wave packet is not smaller than the characteristic gravitational length $l_i$, namely $\sigma \gtrsim l_i$, we can approximate the coefficients with
\begin{equation}
    c_{i,n}^\pm \sim \left(\frac{2\sigma^2}{\pi}\right)^{\frac{1}{4}} \frac{N_{i,n}}{l_{\sigma,i}} \, 
    e^{-\left(\frac{z_n + l_{\pm,i}}{\sqrt{2} l_{\sigma,i}}\right)^2}.
\end{equation}

We now consider the superposition of Gaussian wave packets with internal degrees of freedom of Eq.~\eqref{eq:initial_state_MZ}. In order to be consistent with the choice of the potential above, the wave function of the atom should be null at the floor $x=0$, and in general a Gaussian wave function is not. Therefore, we must require that the initial distance from the floor is large compared to the initial spread of the wave packet, so that the amplitude is negligible at the floor. Thus, we require that
\begin{equation}
    x_\pm \gg \sigma.
\end{equation}
The evolved state is
\begin{equation}\label{eq:FFwF_state}
    \hat U\frac{\ket{\psi_+} + e^{i\varphi} \ket{\psi_-}}{\sqrt{2}}\ket{\tau_{in}} =\sum_{i,n} c_{i,n} e^{-\frac{E_{i,n}}{\hbar} t}\ket{\psi_{i,n}}\ket{E_i}.
\end{equation}

Its QFI in the limit of long times is
\begin{align}
    G^f(g) \sim & \frac{4 \Delta t^2}{\hbar^2} \text{Var}(\partial_g E), \\ 
    \text{Var}(\partial_g E)  = & \sum_{i,n}\ |c_{i,n}|^2 \ (\partial_g E_{i,n})^2 \notag \\ & - \ \left(\sum_{i,n}|c_{i,n}|^2 \ \partial_g E_{i,n}\right)^2.
\end{align}

Neither the coefficients $c_{i,n}$ nor the eigenvalues $E_{i,n}$ depend on time, meaning that the time scaling of this QFI is $\Delta t^2$. Therefore, we obtain that the presence of the floor removes the $\Delta t^4$ and $\Delta t^6$ scaling that was caused by the gravitational time dilation, and also the $\Delta t^4$ dependence that was originated by the vertical motion of the atom. 

These results are consistent with those of  \cite{seveso}, where it was shown that the $\Delta t^2$ scaling in the QFI is a feature of every pure statistical model with discrete eigenvectors, namely such that $\ket{\psi_\lambda} = e^{-i \hat H_\lambda t} \ket{\psi_0}$, where $\ket{\psi_0}\in \mathcal{H}$, $\text{dim}(\mathcal{H})<\infty$.
We expect the floor to have an effect on the dynamics only when its distance from the atom is comparable with the spread of the atomic wave function. Therefore, in order to exploit the higher time scaling of the freely falling interferometer of Section~\ref{s:ffi}, the atom should be revealed far from the floor, where the distance is comparable to the spread of the wave function.

%\printbibliography
\bibliographystyle{quantum}
\bibliography{bibliography}
\end{document}